\documentclass{emulateapj}
\renewcommand{\*}{$^*$}

\newcommand{\uv}{\mbox{$u$-$v$}}
\newcommand{\ex}[1]{\mbox{$\times 10^{#1}$}}

\newcommand{\kms}{\mbox{km s$^{-1}$}}
\newcommand{\muas}{\mbox{$\mu$as}}
\newcommand{\muasyr}{\mbox{$\mu$as yr$^{-1}$}}
\newcommand{\Ra}[4]{\mbox{${#1}^{\rm h} \; {#2}^{\rm m} \; {#3}\fs{#4} $}}
\newcommand{\dec}[4]{\mbox{${#1}\arcdeg \; {#2}\arcmin \; {#3}\farcs{#4} $}}

\shortauthors{Bietenholz et~al.}
\shorttitle{The Location of the Core in M81}

\begin{document}
      
\title{The Location of the Core in M81}

\author{M. F. Bietenholz and N. Bartel}
\affil{Department of Physics and Astronomy, York University, Toronto, M3J~1P3, Ontario, Canada}
 
\author{and M. P. Rupen}
\affil{National Radio Astronomy Observatory, Socorro, New Mexico 87801, USA}

\begin{abstract}

We report on VLBI observations of M81\*, the northwest-southeast
oriented nuclear core-jet source of the spiral galaxy M81, at five
different frequencies between 1.7 and 14.8~GHz.  By phase referencing
to supernova 1993J we can accurately locate the emission region of
M81\* in the galaxy's reference frame.  Although the emission region's
size decreases with increasing frequency while the brightness peak
moves to the southwest, the emission region seems sharply bounded to
the southwest at all frequencies.  We argue that the core must be
located between the brightness peak at our highest frequency
(14.8~GHz) and the sharp bound to the southwest.  This narrowly
constrains the location of the core, or the purported black hole in
the center of the galaxy, to be within a region of $\pm 0.2$~mas or
$\pm 800$~AU (at a distance of $\sim4\rm\,Mpc$).  This range
includes the core position that we determined earlier by finding the
most stationary point in the brightness distribution of M81\* at only
a single frequency.  This independent constraint therefore strongly
confirms our earlier core position.  Our observations also confirm
that M81\* is a core-jet source, with a one-sided jet that extends to
the northeast from the core, on average curved somewhat to the east,
with a radio spectrum that is flat or inverted near the core and steep
at the distant end.  The brightness peak is unambiguously identified
with the variable jet rather than the core, which indicates
limitations in determining the proper motion of nearby galaxies and in
refining the extragalactic reference frame.

\end{abstract}

\keywords{galaxies: individual (M81) --- galaxies: nuclei --- radio continuum: galaxies}

\section{Introduction}

The nearby galaxy M81 (NGC 3031, 0951+693) is a grand-design spiral
which shares some characteristics with radio galaxies and quasars
(Bietenholz et al.\ 1996, Paper~I hereafter).  At a distance of
$\sim$4~Mpc (Freedman et~al.\ 1994; Ferrarese et~al.\ 2000;
Bartel \& Bietenholz 2004), M81 has the distinction of
being the nearest spiral galaxy to have an AGN, with the only
nearer AGN being that in the elliptical galaxy Cen~A (Israel 1998).
The nuclear radio source of M81, which we call M81\*, has an average
brightness temperature of $\sim2\ex{10}$~K at 8.4~GHz
and a slightly inverted synchrotron spectrum, both typical of radio
galaxy cores and quasars.  In particular, its radio spectral index,
$\alpha$, ($S_\nu \propto \nu^\alpha$, where $S$ is the flux density
and $\nu$ the frequency), is about +0.3 from a frequency of
$\sim$1~GHz to its estimated turnover frequency of $\sim$200~GHz
(Reuter \& Lesch 1996).

This is the third paper in a series, in which we continue our
VLBI studies of M81\*\@.  M81\* is extremely compact and is close to the
resolution limit of a ground-based VLBI array at cm wavelengths. We
showed in Paper~I that the size of M81\* along its major axis is only
0.18~mas or 700~AU at a frequency of 22~GHz (see also Bartel et al.\
1982; Kellermann et~al.\ 1976).
It is larger at lower $\nu$ with the length of the major
axis being approximately proportional to $\nu^{-0.8}$.
The orientation also changes, bending from $\sim$40\arcdeg\ at 22~GHz
to $\sim$75\arcdeg\ at 2.3~GHz.  We interpreted this structure as
being due to a core-jet morphology (Paper I).  Even with space
VLBI, the morphology of M81\* was not completely resolved (Bartel \&
Bietenholz 2000).  Only precision astrometry and the determination of
the dynamics and kinematics of different parts of the source allowed
the structure to be unambiguously determined and enabled us to show that M81\*
consists of a stationary core with a one-sided jittery jet to the
northeast (Bietenholz, Bartel \& Rupen 2000, Paper~II hereafter;
also Bietenholz, Bartel \& Rupen 2001),

M81\* also has similarities to the central source in our own Galaxy,
Sgr~A\*, whose radio spectrum is qualitatively very similar, being
slightly inverted up to a turn-over frequency.  Sgr~A\* is much
smaller than M81\* and may represent a version of M81\* scaled down in
power and size. However, Sgr~A\* is largely hidden behind scattering
clouds of gas that have so far allowed only the crudest determinations
of its intrinsic size and orientation (e.g., Bower et~al.\, 2004;
Doeleman et~al.\ 2001).  M81\*, by contrast, is virtually unaffected
by scatter broadening at most radio frequencies and it may therefore
serve as a more readily observable analog to Sgr~A\*.

In this paper, we use multi-frequency astrometry to constrain the
location of the core of M81\*.  In general, identifying the core in a
radio image of an AGN is a challenge.  The most reliable method is to
determine astrometrically which component is stationary relative to a
physically unrelated source that can be assumed to be stationary on
the sky.  This method has previously been used for only two sources,
the superluminal quasar 3C~345 (Bartel et~al.\ 1986) and the double
quasar 1038+528A,~B (Rioja \& Porcas 2000), although it has also been
used to identify a jet component in 4C~39.25 (Guirado et al.\ 1995).

In the absence of kinematic information, the core is usually
identified on the basis of the radio spectrum.  Radio emission from a
core-jet source is synchrotron emission, and most such sources have a
flat or inverted spectrum near the core and steeper spectra farther
along the jet.  Near the core, the optical depth is high because of
synchrotron self-absorption, hence the spectrum tends to being
inverted, while farther along the jet, the optical depth is low and
the spectrum steep.  Another way of identifying the core, then, is by
the region of flattest spectrum.  Such an identification
requires that the images at different frequencies be aligned to an
accuracy better than the resolution.  This can be done if accurate
relative astrometry to a nearby, unrelated source is available, as in
cases like 3C~345 and 1038+528 above.  In the absence of accurate
astrometry, assumptions need to be made, for instance that a
particular peak in the brightness distribution at one frequency
corresponds with that at another.  In some cases, for instance that of
2021+614 (Bartel et~al.\ 1984), the brightness distribution consists
of several distinct components and essentially unambiguous alignment
is possible.  In many other cases, however, the alignment is
problematic, particularly close to the core where a frequency
dependent morphology is expected due to opacity changes (e.g., Lobanov
1998).  In the case of the double quasar 1038+528 a shift in the
position of the brightness peak with frequency was in fact measured
(Marcaide et al.\ 1985).

Like many other AGN, the morphology of M81\* is simple, and separate
components are not clearly discernible.  The alignment of the VLBI
images of M81\* at different frequencies for the purpose of
identifying the regions with flat or inverted spectra is impossible
without accurate astrometrical information.  The coordinates of M81\*,
therefore, need to be accurately determined at different radio
frequencies.  In this paper we report on the location of the radio
emission region in M81\* at five different frequencies spanning an
order of magnitude.  The effective source size decreases with
increasing frequency (Paper I), so at higher frequency, the emission
region is smaller, and hence more accurately constrains the core
location, if we make the reasonable assumption that the core is in
fact within the emission region.  Closely related to this study is the
computation of a spectral index image of M81\*.  However, since M81\*
is rapidly time-variable, we defer the subject of its time-variable
radio spectral index distribution to a forthcoming paper.

The observations of M81\* were
carried out in conjunction with our ongoing VLBI observations of
SN~1993J, for which M81\* is used as a phase reference.  Our first
paper (Paper I) reported on the extremely small size of M81\* and its
orientation, and their frequency dependence. Our second paper (Paper
II) reported on the kinematics and the identification of a stationary
core and a one-sided jittery jet in M81\*. In this paper
we continue our series by using multifrequency astrometry to tightly
constrain the core location in a manner largely independent from the
study in Paper~II\@.

We describe our observations in \S~\ref{sobs}, and show representative
images of M81\* in \S~\ref{simg}. We elaborate on the use of SN~1993J
as an astrometric reference source in \S~\ref{ssn93jref} , and
determine the location of the emission region as a function of
frequency in \S~\ref{spkwave}.  We elaborate on the resulting tight
constraints on the location of the core in \S~\ref{score}, discuss our
results in \S~\ref{sdiscuss}, and give a summary of our conclusions in
\S~\ref{sconc}.

\section{Observations}
\label{sobs}

As mentioned, M81\* was observed as a phase reference source for the
continuing multi-frequency program of VLBI observations of SN~1993J\@.
We repeat a brief description of the observations here for the
convenience of the reader.  A more complete description is given in
Bartel et~al.\ (2002).
We report here on the 32 epochs of VLBI observations completed in 
mid-2003, made with a global array of
between 11 and 18 telescopes with a total time of 12 to 18 hours for
each run, giving us exceptionally good \uv~coverage.  Each telescope
was equipped with a hydrogen maser as a time and frequency
standard. The data were recorded with the VLBA (Very Long Baseline
Array) and either the MKIII or the MKIV VLBI system with sampling
rates of 128 or 256~Mbits per second.  In each session, data at one or
more frequencies of a total of  five (14.8, 8.4, 5.0, 2.3, and 1.7~GHz)
were recorded, with 8.4 and 5.0~GHz being the standard frequencies
used in almost every session.  (There were additional observations 
of SN~1993J but not M81 at 14.8 and 22~GHz, which
are not included in this paper.)  The VLBI data were correlated
using the VLBA processor in Socorro, NM\@.  Further data reduction,
i.e., initial calibration, editing, and fringe-fitting, was done with
NRAO's software package, AIPS.

\section{Images of M81\*}
\label{simg}

\begin{figure}[ht]
\epsscale{0.45}
\plotone{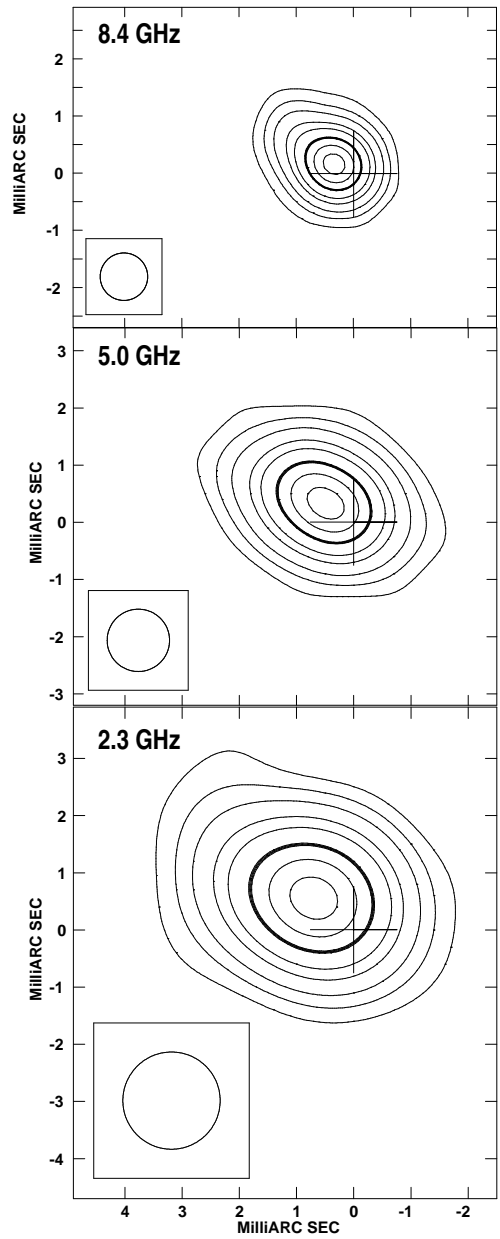}
%C use m81dc96-3panel.fig to make this from AIPS plot files
\figcaption{Images of M81\* on 1996 December 13 at observing
frequencies of 8.4, 5.0 and 2.3~GHz.  In each panel, the origin of the
coordinate system is marked by a cross and is the core position 
derived in Paper II, the FWHM size of the restoring beam is indicated
at lower left, and the contours are drawn at 2, 5, 10, 20, 30, 40,
{\bf 50} (heavy contour), 70 and 90\% of the brightness peak. North is
up and east is to the left.
\label{fimages}}
\end{figure}

In Figure \ref{fimages} we show three representative images of M81\*
taken on 1996 December 13, at frequencies of 8.4, 5, and 2.3~GHz.  At
each frequency, M81\* has a simple brightness distribution, elongated
approximately in the southwest to northeast direction.  The structure
is slightly asymmetric, being somewhat more extended to the
northeast, consistent with our earlier findings in Papers I and II\@.
No sub-components are identifiable in our images.  The structure is,
however, clearly different at the different frequencies.  It can be
seen that the effective size and orientation vary with frequency,
with both the size and position angle, p.a., (north through east)
decreasing with increasing frequency.

\section{The Center of SN 1993J as an Astrometric Reference Point}
\label{ssn93jref}

Phase-referencing allowed us to obtain very accurate relative
positions for M81\* and SN~1993J\@.  The two sources are fortuitously
very close on the sky, with a separation of only $\sim$170\arcsec,
which has the benefit that many systematic sources of error when
determining the relative positions are greatly reduced.  SN~1993J,
however, is not an ideal reference source: it is extended and it
changes with time.  In order for it to serve as our positional
reference, we must select a particular point in its radio morphology
that is stationary on the sky and can be reliably determined at each
epoch and frequency.

Fortunately, the morphology of SN~1993J allows the selection of such a
stationary reference point. Although SN~1993J is expanding rapidly, it
has a well defined shell structure which, in projection, has remained
circular to within 3\%
over the 9 years of observations since the explosion (Bietenholz et
al.\ 2001, 2003; see also Marcaide et al.\ 1997).  This circular
structure suggests isotropic expansion about a stationary center. In
particular, if the geometric center can be shown to be stationary,
i.e., to remain at the position of the explosion center, then it would
be an almost ideal reference point, independent of epoch and
frequency.  We determined the position of the geometric center of
SN~1993J by fitting the projection of a three-dimensional spherical
shell model directly to the \uv~data.  By monitoring the position of
the center of the shell relative to points in the brightness
distribution of M81\* at 8.4~GHz, we found both a nominal proper
motion of the center and the most stable point in the brightness
distribution of M81\* (Paper II; Bietenholz et al.\ 2001).  The
nominal proper motion is small, and thus the geometric center of the
shell virtually coincides with the explosion center, having a proper
motion of $< 20.7$~\muasyr\ or 360~\kms\ ($1\sigma$ upper limit).

Is the position of the geometric center of SN~1993J frequency
dependent?  If the structure of the shell were very different at
different frequencies then the geometric center position might be
frequency dependent.  However, the structure of SN~1993J at 5.0 and
1.7~GHz (Bietenholz et al.\ 2003) and the radius at 15~GHz (see Bartel
et al.\ 2002) are similar to those at 8.4~GHz. Could small differences
between the brightness distributions at the different frequencies
still lead to significant frequency dependence of the position of the
center?  We do not think so: the shell at 8.4~GHz is circular to
within 3\%, and the position of the center is not significantly
affected by brightness asymmetries in the shell (Bietenholz et al.\
2001).  The shell at the other frequencies is similarly circular and
any relatively small differences in the brightness distribution along
the ridge of the shell would have an equally insignificant effect on
the position determination of the center.

We conclude that the geometric center of our shell model of SN~1993J
is stationary within the errors and virtually frequency independent,
and is thus an almost ideal astrometric reference point for M81\*.  We
therefore use the geometric center of the shell of SN~1993J as a
reference point for each epoch and each frequency and determine the
relative coordinates of the images of M81\*, whose structure does show
pronounced and systematic changes with frequency (e.g., Paper~I).  We
adopt R.A. = \Ra{09}{55}{24}{7747593}, decl.\ =
\dec{69}{01}{13}{703188} (J2000) for the coordinates of the geometric
center of SN~1993J at each epoch.  These coordinates are
those of the explosion center of SN~1993J (see Bietenholz et~al.\ 2001) and
agree with the International Celestial Reference Frame (ICRF)
coordinates of SN~1993J (IERS 1999) within the standard errors of 0.6~mas of 
the latter.  We note however, that our results in the present paper do not
depend significantly on any particular choice of the ``absolute''
position values, but only on the relative positions of M81\* and
SN~1993J.

We also note that the geometric center of SN~1993J may not be quite
stationary, since we did find a nominal proper motion with respect to
the core of M81\*\ of $11.4 \pm 9.3$~\muasyr\ ($200 \pm 160$~\kms,
Bietenholz et al.\ 2001).
Does this motion need to be taken into account?  If the proper motion
is really that of the center of SN~1993J, whether due to the galactic
rotation of SN~1993J, the motion of SN 1993J's progenitor, or to
anisotropy in the expansion, it would affect our results on the
relative position determinations of M81\*, since we would then be
using a moving reference point.  However, since the nominal proper
motion is small, being equivalent to a displacement of only 0.10
($\pm 0.08$)~mas over our nine-year observing period, the effect would be
small also.  Further, since our present results are based on
time-averages, any effect of the nominal proper motion of SN~1993J's
geometric center would be even less than 0.10~mas and thus not
significant.

\section{The Radio Emission Region as a Function of Frequency}
\label{spkwave}

\begin{figure}
\epsscale{0.8}
\plotone{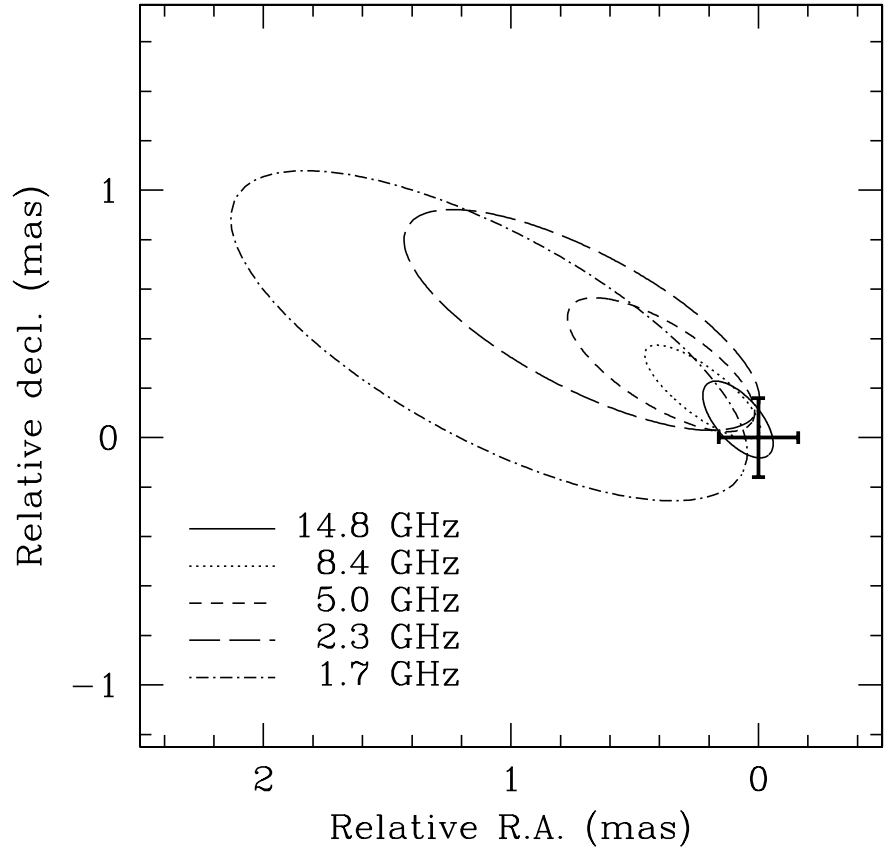}
%C use m81_posnfreq.sm v2.8 to generate this figure (run
%C  subroutine ellpanel in freq not wavelength)
\figcaption{The 50\% contour of the average over all epochs of the
elliptical Gaussians fit to the emission at each frequency.  The
average elliptical Gaussian at each frequency is formed by averaging
separately the R.A. and decl.\ offsets, the FWHM major and minor axes,
and the p.a.\ of the fitted Gaussians at each epoch. The center position
of the Gaussians is determined with respect to the geometric center of
SN~1993J\@.  The origin of the coordinate system is the core position
determined in Paper~II and Bietenholz et al.\ (2001), which is at R.A. = 
\Ra{09}{55}{33}{173063},
decl.\ = \dec{69}{03}{55}{061464} (J2000), and is indicated, along with
its uncertainty, by the cross.
\label{fellwave}}
\end{figure}

Armed with the geometric center of SN 1993J as an astrometric
reference point, we can proceed to accurately determine the position
of points in the brightness distribution of M81\*.  First we examine
how the emission region of M81\* varies as a function of frequency.
Since the source is only partially resolved, we characterized the
emission region by fitting elliptical Gaussians, with the fitting done
in the \uv~plane.  We fitted elliptical Gaussians at each epoch and
frequency.  The center of the fitted Gaussian is virtually coincident with the peak
brightness point discussed below\footnote{While the center of the fitted
Gaussian is not formally identical to the peak brightness point of an
arbitrary brightness distribution, they are coincident to within a
small fraction of the positional uncertainties for our data.}.  At each
frequency, the fitted Gaussian varies with epoch, with the fractional rms
variation of the major axis being $\sim 25$\%, and the rms variation
in the p.a.\ being $\sim 10$\arcdeg.  We discussed the temporal
variability at 8.4~GHz in Paper~II, and we defer a detailed
discussion of the temporal variability at other $\nu$ to a future
paper.  For our present purposes, we only require an overall
description of the emission region, and a more complicated model, such
as that used in Paper~II, is not necessary.  In Figure~\ref{fellwave}
we plot the 50\% contour of the average elliptical Gaussian at
each frequency.  It can be seen that the southwest extremity of the
50\% contour is roughly coincident at all frequencies, while the
northeast one varies considerably, being $\sim$2~mas farther to the
northeast at 1.7~GHz than at 14.8~GHz.

The correspondence of the southwest 50\% points of the average fitted
Gaussians at the different frequencies suggests that the actual
emission region has a sharp boundary to the southwest which is
independent of frequency.  A one-sided jet is suggested, with the core
being near this 50\% point, and the jet extending to the northeast.
The extent of the emission to the northeast, or the length of the jet,
is roughly proportional to $\nu^{-0.8}$, consistent with our
findings in Paper~I.

How does the location of the brightness peak vary with frequency?  In
Table~\ref{tpeakpos} we give the positions of the brightness peak of
M81\* with respect to the position of the center of SN~1993J's shell.
For easier reading, we tabulate the positions as offsets from the
position of the core of M81\* derived in Bietenholz et al.\ (2001),
namely R.A. =
\Ra{09}{55}{33}{173063} and decl. = \dec{69}{03}{55}{061464} (J2000).

We adopt constant standard errors at each frequency.  A detailed
discussion of the astrometric uncertainties at 8.4~GHz can be found
in Paper~II and Bietenholz et al.\ (2001), and at that frequency we
adopt the standard error of 40~\muas\ used there.  The uncertainties
at our other frequencies are scaled with $\nu$ from the value at
8.4~GHz, with the exception of those at 14.8~GHz, for which a somewhat
higher value is taken because of the generally smaller VLBI arrays and
thus poorer \uv~coverage at that frequency.

We plot the position of the brightness peak at each epoch and
frequency in Figure~\ref{fcorepos}.  The position of the brightness
peak in M81\* is somewhat variable at each frequency, as we had
already found for 8.4~GHz in Paper~II\@.  However, a clear picture
emerges when all the points are considered.  At each frequency, we
determined the average position of
the brightness peak.  These average positions are also plotted in
Figure~\ref{fcorepos} and listed in Table~\ref{tavgpos}.  It is clear
that the average position of the brightness peak is a well-behaved
function of $\nu$, lying farther to the northeast for lower $\nu$ (see
Ebbers et al.\ 1998 for a preliminary version of this result).  The
positions of the brightness peak as a function of $\nu$ lie along a
curved line on the sky, with the p.a.\ of the line being
$\sim$55\arcdeg\ at the higher frequencies, and changing to
$\sim$90\arcdeg\ by 1.7~GHz.
In other words, the emission region in M81\* is, on average, bent
somewhat to the east, with the emission at high frequencies coming
predominately from the southwest.  This is consistent with the change
in p.a.\ with frequency observed earlier (Bartel et~al.\ 1982; Paper~I),
but gives a clearer picture of the jet geometry.

\section{The Location of the Core}
\label{score}

We have shown that the emission region seems to be sharply bounded,
independent of frequency, to the southwest, and that both the size of
the emission region and the position of the brightness peak are
functions of frequency, with the size increasing and the 
brightness peak moving to the northeast with decreasing frequency.

In Papers I and II, we showed that M81\* could be interpreted as a
source with a core and a one-sided jet.  In such sources, the radio
spectrum is typically flat or inverted near the core where the source
is optically thick due to synchrotron self-absorption, and steep
farther along the jet where the source becomes optically thin.  Such a
source produces precisely the frequency-dependent morphology seen in
M81\*, with the brightness peak moving closer to the core with
increasing frequency.  The coincidence of the southwest 50\% points at
different frequencies suggests that any counterjet is indeed faint in comparison to
the jet, and thus that the core is near the southwest 50\% point.  By
assuming that the core is within, or at least on the edge of the
emission region, we can then use the emission region at our highest
frequency of 14.8~GHz, which is the smallest and the most accurately
located, to set a fairly small range of possible core locations.

\begin{figure}[ht]
\epsscale{0.8}
\plotone{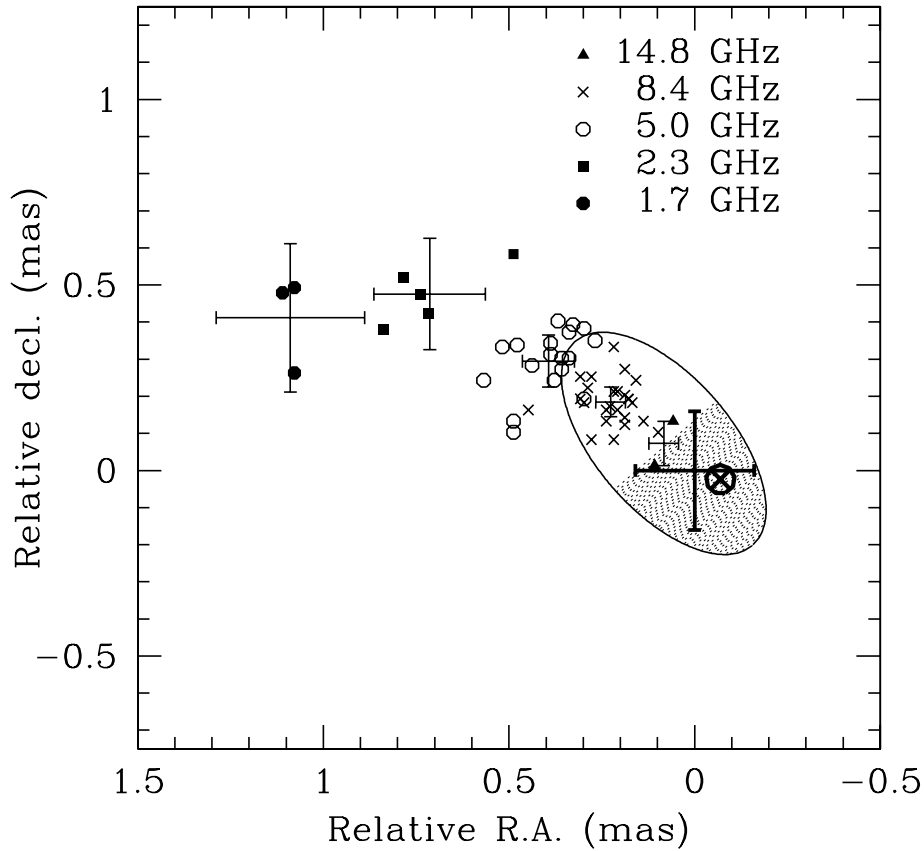}
%C  use m81_posnfreq.sm (v2.8) to generate this figure 
\figcaption{The position of the brightness peak of M81\* with respect
to the geometric center SN~1993J for each epoch of observation.  The
points show the positions of the peak at each epoch at the frequencies
indicated.  The estimated standard errors of each measurement in
R.A. and decl.\ are 40, 40, 70, 150 and 200~\muas\ at frequencies
of 14.8, 8.4, 5.0, 2.3 and 1.7~GHz respectively. 
The thin crosses show the average positions over all epochs at each
frequency, with the plotted standard error being the larger of the rms
variation in the average position and the positional uncertainty of
each individual measurement at that frequency.  The heavy cross shows
the core position of R.A. = \Ra{09}{55}{33}{173063}, decl.\ =
\dec{69}{03}{55}{061464} (J2000; Paper~II and Bietenholz et~al.\
2001), derived as the most stationary point in the brightness
distribution of M81\* at 8.4~GHz, which we also use as the origin of
our coordinate system, with its standard error.  The ellipse shows the
average extent of the emission at 14.8~GHz or $\lambda = 2.0$~cm (10\%
contour), with the shaded half indicating a conservative estimate of
the range of possible core locations (\S~\ref{score}).  The
$\bigotimes$ shows an extrapolation of the position of the brightness
peak to $\lambda = 0$ (see \S~\ref{score}).
\label{fcorepos}}
\end{figure}

As a conservative estimate of the emission region at 14.8~GHz,
we take the 10\% contour (rather than the 50\% contour plotted
earlier) of the average fitted Gaussian to delimit the emission region
at that wavelength.  That average fitted Gaussian\footnote{The
uncertainties given are standard deviations over five observing
epochs, representing both the measurement uncertainties and intrinsic
variation of the size and orientation of the radio source.} had a
major axis of $0.4\pm0.1$~mas, a minor axis of $0.2^{+0.1}_{-0.2}$~mas
(both FWHM) and p.a.\ of $41 \pm 13$\arcdeg.
The 10\% contour of that Gaussian is also plotted in
Figure~\ref{fcorepos}.

Since the size of the emission region decreases and the position of
the brightness peak moves to the southwest with increasing frequency
over our entire 10 to 1 frequency range, it seems reasonable to
assume that this behavior would continue to even higher frequencies,
perhaps even to $\sim 200$~GHz where the integrated spectrum
begins to steepen (Reuter \& Lesch 1996).  If we assume that the core
is located within the emission region, and near the brightness peak,
at such high frequencies, then the core location must be to the
southwest of the brightness peak even at our highest frequency of
14.8~GHz.

Combining these two constraints, namely that the core be within the
emission region at 14.8~GHz and that it be southwest of the
brightness peak at that same frequency, allows us to tightly
constrain the region where the core could be located, and we indicate
this region by the
shaded half-ellipse drawn in Figure~\ref{fcorepos}.  Our earlier
determination of the core position in Paper~II and Bietenholz et al.\
(2001), which is also plotted in Figure~\ref{fcorepos}, falls very
near the center of this region of possible core locations.

Since the location of brightness peak seems to be a well-behaved
function of frequency, or correspondingly of the observing wavelength,
$\lambda$, one could extrapolate this function to $\lambda = 0$, or
perhaps at least to $\lambda \sim 1$~mm (i.e., $\nu \sim 200$~GHz)
where the integrated spectrum turns over, and use the extrapolated
position as an estimate of the core location.  Such an extrapolation,
however, requires knowledge of the functional form of the dependence
of the position on $\lambda$.  Neither the physics of the radio emission
nor the jet geometry are well known enough to determine a
particular choice for this function.  As an illustrative example, we
chose a simple linear function of $\lambda$, which we fit by weighted
least-squares to both the R.A. and the decl.\ of the brightness peak
for our three shortest wavelengths, those being the ones closest to
the core and therefore providing the most useful constraints on the
core location.  Since only a relatively small extrapolation is
required between our shortest observed wavelength of $\lambda =
2.0$~cm and $\lambda = 0$, we think that our extrapolation is
reasonable despite being non-unique.  The intercept of these fits give
us a position at $\lambda = 0$ of R.A. = \Ra{09}{55}{33}{173050} and
decl.\ = \dec{69}{03}{55}{061440} (J2000),
which we also plot in Figure~\ref{fcorepos} ($\bigotimes$).  This
position is less than 75~\muas\ away from the core position derived in
Paper~II and Bietenholz et al.\ (2001), and thus well 
within the uncertainty of 160~\muas\ of the latter.

\section{Discussion}
\label{sdiscuss}

Using VLBI observations of M81\* at 32 epochs and at a total of five
frequencies, all phase-referenced to a physically unrelated source in
the same galaxy (SN~1993J), has allowed us to place tight constraints
on the location of the core, showing it must be within a region of
about $\pm 0.2$~mas of the southwestern part of the brightness
distribution at our highest frequency of 14.8~GHz.  This is consistent
with the position of the core we derived earlier by finding the most
stable point within the radio morphology of M81\* (Paper~II).
Furthermore the constraints on the core location are of comparable
size to the uncertainties of the earlier core position.

The present finding provides an important independent confirmation of
our earlier core location.  Although the determination of the core
position from Paper~II was obtained from a subset of the observations
used for the present one, our present results are largely independent
of the earlier core position, since that earlier position was based on the
time-variability of the morphology of M81\* at only one frequency.
The present results, by contrast, are based upon a time-average of the
peak position at different frequencies.  Any possible errors in the
astrometry would seem unlikely to affect the two determinations of the
core position in the same manner.

The positions of the brightness peak at different frequencies, plotted in
Figure~\ref{fcorepos}, also draw a clear picture of the average jet
geometry, with the jet extending to the northeast of the core, and
being bent somewhat to the east.  This bending is consistent with what
was found in Bartel et~al.\ (1982) and Paper~I\@.  However, we showed
in Paper~II that the geometry of M81\*, at least at 8.4~GHz, was
intrinsically variable: between 1993 and 1998, the FWHM major axis
size varied between 0.40 and 0.70~mas and its p.a.\ between 44\arcdeg\
and 57\arcdeg.  This variability is reflected in the scatter of the
brightness peak positions at each frequency in Figure~\ref{fcorepos}.
This scatter does not affect our core location, since the core can be
assumed stationary, and our core location is derived from the
time-averaged positions at each frequency.  In any case, the scatter is
considerably smaller than the variation of the peak position with
frequency from which our core location is derived.

In general, the identification of the core will be important for
unambiguously determining the spectral index distribution of M81\*,
for instance to determine the spectral index distribution of the jet.
Further, the identification of the core, as we noted in Paper~II, is
important for efforts to determine the proper motion of the galaxy M81
in the extragalactic reference frame.

Finally the independent confirmation of our earlier core position to
within its uncertainties strongly supports our identification of the
explosion center of SN~1993J, the small limits on the proper motion of
SN1993J's geometric center, and the near isotropy of its expansion
(Bietenholz et al.\ 2001, 2003).

\subsection{The Spectral Index Distribution of M81\*}

The results described in this paper imply a strong gradient in the
radio spectral index of M81\*, since if the spectral index were
uniform, then the morphology would be similar at different frequencies
and no displacement of the brightness peak as a function of frequency
would be seen.  The gradient is such that the spectrum near the
core is flatter than that at the far end of the jet. 
We estimate that $\alpha \gtrsim 0.0$ near the core, and $\alpha \sim
-0.7$ along the jet $\sim 1$~mas away from the core. We will discuss
the spectral index distribution of M81\* in detail in a future paper.

\subsection{The Extragalactic Reference Frame}
\label{srefframe}

In general, our results indicate limitations to the accuracy of the
extragalactic reference frame as determined with VLBI observations.
Most of the sources defining the ICRF extragalactic reference frame
(Ma et al.\ 1998; IERS 1999) have core-jet structure, which often
shows some variability with time (Fey, Clegg, \& Fomalont 1996; Fey \&
Charlot 1997, 2000; Gontier et al.\ 2001), due presumably to a
variable jet.  The most accurate determination of the extragalactic
reference frame would therefore be obtained by determining the
positions of the cores of the reference frame sources.  Our results
show that, even in a source as compact as M81\*, not only does the
peak of the brightness distribution not coincide with the core, but
also that the position of the peak of the brightness distribution is
frequency dependent\footnote{In fact, Da Silva Neto et~al.\ 2002,
reported on discrepancies of up to 15~mas between the radio and the
optical positions of the ICRF reference frame sources, although it is
not clear these discrepancies can be directly attributed to radio
source structure.}.

We note also that an error is made when the effect of the ionospheric
delay is computed from simultaneous observations at 8.4 and 2.3~GHz
with the assumption that the sources' positions coincide at the two
frequencies.  Any discrepancy in the position causes an astrometric
error of about 7\% of the discrepancy.  In the case of M81\* this
error would be about 0.05~mas at 8.4~GHz (we note that it was not
necessary to determine the ionospheric delay in this fashion for our
relative astrometry, so our position values are unaffected by this
error).

However, if the position of the peak in the brightness distribution were
stable over time, then it could be taken as an astrometric fixed point
for the extragalactic reference frame, without having to locate the
core within the brightness distribution.  Our finding that the peak
position is variable by $\sim$0.3~mas at all our frequencies shows
that using the brightness peak (or the centroid of the emission
region) as a reference point limits the positional accuracy to that
level.  In the case of M81\*, the variability in position is only
about one third of the error of the ICRF position. The errors of the
positions of some ICRF sources, however, are as small as 0.1~mas.  The
angular scales on which core-jet structure is seen or expected for the
ICRF sources is from 0.1~mas upward.  M81\* is a very compact source,
and since the position of the peak of the brightness distribution is
variable on a scale of 0.3~mas it is conceivable that the stability of
the positions of even the most compact ICRF sources is limited to about
$\pm 0.1$~mas.

\subsection{The Proper Motion of M81} 

The dynamics of galaxies of the Local Group and beyond are an important
area of research in astronomy.  Since a proper motion has not yet been
measured for any galaxy, studies of the galaxy dynamics are based
exclusively on the radial component of their velocities. The galaxy
M81 is one of the few whose proper motion could perhaps be measured in
the not so distant future.  Observations of the core in M81\*
relative to quasars nearby on the sky over a period of about 10 years
might give the proper motion of the galaxy with a standard
error $\lesssim 100$~\kms, since unlike M81\*, the cores of the more
distant quasars are not expected to display any discernible proper motion
(see e.g., Bartel et al.\ 1986).   In fact, since Nagar et~al.\
(2002) have shown that many nearby low-level AGNs have mas-scale
cores with brightness temperatures $> 10^8$~K, such analysis could
be extended to other nearby galaxies.

However, as we point out in \S~\ref{srefframe} above, the intrinsic
variability of core-jet sources, including M81\* itself, limits the
accuracy with which the proper motion of M81\* could be determined in
this fashion. For the highest accuracy, therefore, one would need repeated
observations of M81\* together with several reference sources,
and an analysis similar to the one we carried out in Paper~II and
Bietenholz et al.\ (2001) to locate the cores as
most stable points within the brightness distributions of each
reference source.

\section{Conclusions}
\label{sconc}

\begin{trivlist}

\item{1.} We determined the size of the emission region
of M81\* and its location with respect to the geometric center of
SN~1993J at five different observing frequencies spanning an order
of magnitude.

\item{2.} The size of the emission region decreases with increasing
observing frequency.  The position of the brightness
peak shifts progressively toward the southwest extremity with
increasing frequency.  The position of that southwest
extremity is largely frequency-independent, suggesting a hard edge to
the emission region at that point. 

\item{3.} This progressive shift is expected for the emission from a
compact core-jet source with a flat or inverted spectrum near the
core and a steep spectrum at the far end of the jet, and thus confirms
our earlier identification of M81\* as a core-jet source.

\item{4.} We argue that the core is within the southwest half of the
emission region at our highest frequency of 14.8~GHz, which constrains
the core to be located within a region of about $\pm 0.2$~mas.  This
constraint is consistent with, but independent from, our earlier
determination of the core position, and thus strongly confirms the
latter.

\item{5.} The confirmation of the core location in M81\* also confirms
our identification of the explosion center of SN~1993J and its
isotropic expansion therefrom.

\item{6.} The brightness peak of M81\* is unambiguously identified
with the jet rather than with the core.  The location of the
brightness peak relative to the stationary core is variable both in
time and with frequency.

\item{7.} An accurate location of the core, as opposed to the
brightness peak, will be important in determining the proper motion of
nearby galaxies in particular, and in refining the extragalactic
reference frame in general.
\end{trivlist}

\acknowledgements

We thank V. I. Altunin, A. J. Beasley, W. H. Cannon, J. E. Conway,
D. A. Graham, D. L. Jones, A. Rius, G. Umana, and T. Venturi for help
with several aspects of the project.  R. Bartel, J. Cadieux, M. Craig,
R. Freedman, M. Keleman, and B. Sorathia helped with some aspects of
the VLBI data reduction during their tenure as students at York. We
thank NRAO, the European VLBI Network, and the NASA/JPL Deep Space
Network (DSN) for providing exceptional support for this extended and
ongoing observing campaign. We also thank Natural Resources Canada for
helping with the observations at the Algonquin Radio Observatory
during the first years of the program.  Research at York University
was partly supported by NSERC\@.  NRAO is operated under license by
Associated Universities, Inc., under cooperative agreement with NSF\@.
The European VLBI Network is a joint facility of European and Chinese
radio astronomy institutes funded by their national research councils.
The NASA/JPL DSN is operated by JPL/Caltech, under contract with
NASA\@.  We have made use of NASA's Astrophysics Data System Abstract
Service.

\newcommand{\xsp}{\hspace{0.4in}}
\newcommand{\Da}{$\Delta\alpha$}
\newcommand{\Dd}{$\Delta\delta$}
\newcommand{\cdt}{\nodata}
\begin{deluxetable}{l c c @{\xsp} c c @{\xsp} c c @{\xsp} c c @{\xsp} c c}
\tabletypesize{\footnotesize}
\tablecaption{Position Offsets of the Brightness Peak of M81\*}
\tablehead{
\multicolumn{1}{c}{Date} & \multicolumn{10}{c}{Position Offset} \\
& \multicolumn{10}{c}{(mas)}  \\  \cline{2-11}
& \multicolumn{2}{c @{\xsp}}{14.8~GHz} 
& \multicolumn{2}{c @{\xsp}}{8.4~GHz}
& \multicolumn{2}{c @{\xsp}}{5.0~GHz}
& \multicolumn{2}{c @{\xsp}}{2.3~GHz}
& \multicolumn{2}{c @{\xsp}}{1.7~GHz} \\
           &\Da\tablenotemark{a}&\Dd\tablenotemark{a}
           &\Da&\Dd&\Da&\Dd&\Da&\Dd&\Da&\Dd 
}
\startdata
1993 May 17 & 0.06 & 0.13 & 0.10 & 0.10 & \cdt & \cdt & \cdt & \cdt & \cdt & \cdt \\
1993 Jun 27 & \cdt & \cdt & 0.19 & 0.12 & \cdt & \cdt & \cdt & \cdt & \cdt & \cdt \\
1993 Sep 19 & 0.11 & 0.01 & 0.14 & 0.13 & \cdt & \cdt & \cdt & \cdt & \cdt & \cdt \\
1993 Nov  6 & \cdt & \cdt & 0.16 & 0.24 & \cdt & \cdt & \cdt & \cdt & \cdt & \cdt \\
1993 Dec 17 & \cdt & \cdt & 0.28 & 0.08 & \cdt & \cdt & \cdt & \cdt & \cdt & \cdt \\
1994 Jan 28 & \cdt & \cdt & 0.18 & 0.19 & \cdt & \cdt & \cdt & \cdt & \cdt & \cdt \\
1994 Mar 15 & \cdt & \cdt & 0.19 & 0.27 & \cdt & \cdt & \cdt & \cdt & 1.11 & 0.48 \\
1994 Apr 22 & \cdt & \cdt & 0.22 & 0.33 & 0.30 & 0.38 & \cdt & \cdt & \cdt & \cdt \\
1994 Jun 22 & \cdt & \cdt & 0.21 & 0.21 & 0.33 & 0.39 & \cdt & \cdt & \cdt & \cdt \\
1994 Aug 30 & \cdt & \cdt & 0.19 & 0.20 & 0.34 & 0.37 & \cdt & \cdt & \cdt & \cdt \\
1994 Oct 31 & \cdt & \cdt & 0.17 & 0.18 & 0.27 & 0.35 & \cdt & \cdt & \cdt & \cdt \\
1994 Dec 23 & \cdt & \cdt & 0.18 & 0.19 & 0.37 & 0.40 & 0.47 & 0.40 & \cdt & \cdt \\
1995 Feb 12 & \cdt & \cdt & 0.21 & 0.16 & \cdt & \cdt & \cdt & \cdt & \cdt & \cdt \\
1995 May 11 & \cdt & \cdt & 0.24 & 0.16 & 0.36 & 0.30 & \cdt & \cdt & \cdt & \cdt \\
1995 Aug 18 & \cdt & \cdt & 0.28 & 0.25 & 0.39 & 0.34 & \cdt & \cdt & \cdt & \cdt \\
1995 Dec 19 & \cdt & \cdt & 0.24 & 0.13 & 0.44 & 0.28 & 0.84 & 0.38 & \cdt & \cdt \\
1996 Apr  8 & \cdt & \cdt & 0.30 & 0.18 & 0.38 & 0.24 & \cdt & \cdt & \cdt & \cdt \\
1996 Sep  1 & \cdt & \cdt & 0.19 & 0.14 & 0.36 & 0.27 & \cdt & \cdt & \cdt & \cdt \\
1996 Dec 13 & \cdt & \cdt & 0.31 & 0.19 & 0.48 & 0.34 & 0.79 & 0.52 & \cdt & \cdt \\
1997 Jun  7 & \cdt & \cdt & 0.23 & 0.17 & \cdt & \cdt & \cdt & \cdt & \cdt & \cdt \\
1997 Nov 15 & \cdt & \cdt & 0.22 & 0.21 & 0.30 & 0.19 & 0.72 & 0.42 & \cdt & \cdt \\
1998 Jun  3 & \cdt & \cdt & 0.31 & 0.25 & 0.39 & 0.31 & \cdt & \cdt & \cdt & \cdt \\
1998 Nov 20 & \cdt & \cdt & \cdt & \cdt & 0.52 & 0.33 & \cdt & \cdt & \cdt & \cdt \\
1998 Dec  7 & \cdt & \cdt & 0.29 & 0.22 & \cdt & \cdt & 0.74 & 0.47 & \cdt & \cdt \\
1999 Jun  6 & \cdt & \cdt & \cdt & \cdt & \cdt & \cdt & \cdt & \cdt & 1.08 & 0.49 \\
1999 Jun 16 & \cdt & \cdt & \cdt & \cdt & 0.34 & 0.30 & \cdt & \cdt & \cdt & \cdt \\
1999 Nov 24 & \cdt & \cdt & \cdt & \cdt & 0.57 & 0.24 & \cdt & \cdt & \cdt & \cdt \\
2000 Feb 25 & \cdt & \cdt & 0.45 & 0.16 & \cdt & \cdt & \cdt & \cdt & \cdt & \cdt \\
2000 Nov 13 & \cdt & \cdt & 0.22 & 0.08 & \cdt & \cdt & \cdt & \cdt & \cdt & \cdt \\
2001 Jun 10 & \cdt & \cdt & \cdt & \cdt & 0.49 & 0.13 & \cdt & \cdt & \cdt & \cdt \\
2001 Nov 26 & \cdt & \cdt & \cdt & \cdt & \cdt & \cdt & \cdt & \cdt & 1.08 & 0.26 \\
2002 May 25 & \cdt & \cdt & \cdt & \cdt & 0.49 & 0.10 & \cdt & \cdt & \cdt & \cdt \\
\colrule
Standard Error:\tablenotemark{b} 
            & 0.04 & 0.04 & 0.04 & 0.04 & 0.07 & 0.07 & 0.15 & 0.15 & 0.20 & 0.20 \\
\enddata
\tablenotetext{a}{The R.A.\ ($\alpha$) and decl.\ ($\delta$) offsets
of the position of the brightness peak of M81\* from the core position of
\Ra{9}{55}{33}{173063}, \dec{69}{3}{55}{061464} derived in Bietenholz et al.\
(2001) at 8.4~GHz, determined from astrometry to SN~1993J and taking the center
position of SN~1993J to be \Ra{9}{55}{24}{7747593},
\dec{69}{1}{13}{703188} (J2000)}
\tablenotetext{b}{Standard errors, see text \S\ref{spkwave}}
\label{tpeakpos}
\end{deluxetable}

\begin{deluxetable}{c @{\xsp} c @{\hspace{1in}} c c @{\hspace{1in}} c c}
\tablecaption{Average Position of the Brightness Peak of M81\* as a Function 
of Frequency}
\tablehead{
Frequency  & $N_{\rm obs}\tablenotemark{a}$ 
 & \multicolumn{4}{c}{Average Position Offset\tablenotemark{b}} \\
(GHz)       &         & \multicolumn{4}{c}{(mas)} \\ \cline{3-6}
           &         & \Da & rms$_\alpha$ & \Dd & rms$_\delta$
}
\startdata
    14.8 & \phn2 & 0.08 & 0.04 & 0.07 & 0.08 \\
\phn 8.4 &    25 & 0.23 & 0.07 & 0.18 & 0.06 \\
\phn 5.0 &    18 & 0.39 & 0.08 & 0.29 & 0.09 \\
\phn 2.3 & \phn5 & 0.71 & 0.13 & 0.48 & 0.08 \\
\phn 1.7 & \phn3 & 1.09 & 0.02 & 0.41 & 0.13 \\
\enddata  
\tablenotetext{a}{The number of observing epochs at each frequency}
\tablenotetext{b}{The average over all observing epochs of the R.A.\
($\alpha$) and decl.\ ($\delta$) offsets of the position of the
brightness peak of M81\* and their rms variation over all of our
observing epochs.  The offsets in position are from
\Ra{09}{55}{33}{173063}, \dec{69}{03}{55}{061464} (J2000),
see Table~\ref{tpeakpos}}
\label{tavgpos}
\end{deluxetable}

\end{document}